\documentclass[amsmath, amsthm, amssymb, floatfix, superscriptaddress, twocolumn, aps, 10pt]{revtex4-2}
\usepackage[hidelinks]{hyperref}
\usepackage{graphicx}
\usepackage{booktabs} 
\usepackage{array} 
\usepackage{tablefootnote}
\usepackage{threeparttable}

\usepackage[labelfont=bf, justification=raggedright, figurename={Fig.}, tablename={Table}, labelsep=period, font=small]{caption}

\graphicspath{{figure/}}

\begin{document}
\title{Subvolt high-speed free-space modulator with electro-optic metasurface}
\author{Go~Soma}
\email{go.soma@tlab.t.u-tokyo.ac.jp}
\affiliation{School of Engineering, The University of Tokyo, Tokyo, Japan}
\author{Koto~Ariu}
\affiliation{School of Engineering, The University of Tokyo, Tokyo, Japan}
\author{Seidai~Karakida}
\affiliation{School of Engineering, The University of Tokyo, Tokyo, Japan}
\author{Yusuke~Tsubai}
\affiliation{School of Engineering, The University of Tokyo, Tokyo, Japan}
\author{Takuo~Tanemura}
\email{tanemura@ee.t.u-tokyo.ac.jp}
\affiliation{School of Engineering, The University of Tokyo, Tokyo, Japan}

\begin{abstract} 

\noindent
Active metasurfaces incorporating electro-optic (EO) materials enable high-speed free-space optical modulators that show great promise for a wide range of emerging applications, including free-space optical communication, light detection and ranging, and optical computing.
However, the limited light-matter interaction lengths in ultrathin metasurfaces typically require high driving voltages exceeding tens of volts.
Here we present ultralow-voltage, high-speed free-space optical modulators based on silicon-organic-hybrid metasurfaces with dimerized-grating-based nanostructures. By exploiting a high-\textit{Q} quasi-bound state in the continuum, normally incident light is effectively trapped within a submicrometer-scale silicon slot region embedded with organic EO material.
Consequently, highly efficient modulation is obtained, enabling data transmission at 50~Mbps and 1.6~Gbps with driving voltages of only 0.2~V and 1~V, respectively.
These unprecedented metasurface modulators operating at complementary metal-oxide-semiconductor (CMOS)-compatible voltage levels provide the pathway toward energy-efficient high-speed active metasurface devices.
\end{abstract}
\maketitle

\noindent
Metasurfaces are flat optical components composed of dense arrays of ultrathin subwavelength structures, which can manipulate the intensity, phase, and polarization of incident free-space light \cite{Yu2011-ey, Lin2014-fb, Khorasaninejad2017-sb, Arbabi2022-jk, Kuznetsov2024-lv}. 
With the unique and unprecedented functionalities beyond the capabilities of conventional optics, various types of metasurface-based optical devices have been demonstrated for imaging \cite{Chen2018-lu, Wang2018-cm, Rubin2019-tb, Miyata2021-cg, Fan2024-ih}, holography \cite{Ni2013-be, Balthasar_Mueller2017-ex, Xiong2023-mk, Gopakumar2024-ad}, and optical communication \cite{Oh2022-tp, Soma2023-fx, Komatsu2024-pi}. 
However, many of these previous implementations have been passive, meaning that their functionalities were fixed after fabrication. 
In recent years, there has been a growing interest in developing active metasurfaces that can be reconfigurable \cite{Wang2016-ca, Li2019-ro, Holsteen2019-sn, Zhang2021-mx, Wang2021-od, Shaltout2019-bw, Gu2022-ky}.  
In particular, high-speed metasurface modulators are attractive for a wide range of emerging applications, including free-space optical (FSO) communication, light detection and ranging (LiDAR), and optical computing \cite{Yao2014-aw, Huang2016-wn, Park2017-oq, Wu2019-pz, Shirmanesh2020-yf, Park2021-qt, Panuski2022-zm, Sisler2024-pw}.

Among various approaches, electro-optic (EO) metasurface modulators using $\chi^{(2)}$ nonlinear optical materials, such as organic EO (OEO) materials \cite{Ren2015-qe, Zhang2017-rq, Miyano2022-uc, Sun2022-wt, Zhang2023-gb, Zhang2023-de, Benea-Chelmus2021-re, Sun2021-se, Sun2022-as, Benea-Chelmus2022-di, Ogasawara2019-cp, Zheng2024-xh, Kosugi2016-yg, Fukui2022-my}, LiNbO$_3$ \cite{Weiss2022-vn, Damgaard-Carstensen2022-yl, Damgaard-Carstensen2023-ag, Damgaard-Carstensen2024-rh, Weigand2021-lu, Trajtenberg-Mills2024-sx, Di-Francescantonio2024-cu, Ju2024-qc}, and BaTiO$_3$ \cite{Karvounis2020-qm, Li2024-mk, Weigand2024-vi}, offer the potential for ultra-high-speed and low-loss optical modulation.
High-speed EO metasurface modulators, however, suffer from relatively small refractive index change of EO materials (typically $\Delta n<0.01$) and inherent short light-matter interaction length of surface-normal configuration (unlike waveguide-based modulators), which limit the overall modulation efficiency.
To cope with these issues, highly confined optical resonances with sufficiently high quality (\textit{Q}) factors are generally required.
One approach is to employ nanometallic resonators that offer tight optical confinement and large modal overlap with the applied electrical field \cite{Damgaard-Carstensen2022-yl, Damgaard-Carstensen2023-ag, Damgaard-Carstensen2024-rh, Weiss2022-vn, Zhang2023-gb, Ren2015-qe, Zhang2017-rq, Miyano2022-uc, Sun2022-wt}.
However, the inherent plasmonic loss of the metals inevitably leads to degraded \textit{Q} factors, limiting the modulation efficiency. 
In contrast, dielectric resonating nanostructures that can minimize the optical loss are advantageous in achieving high-\textit{Q} resonances \cite{Kosugi2016-yg, Ogasawara2019-cp, Fukui2022-my, Benea-Chelmus2022-di, Zheng2024-xh}. 
However, previous demonstrations of dielectric-based metasurface modulators suffered from insufficient modal confinement and/or large separation between the electrodes that increased the required voltage.
As a result, EO metasurface modulators demonstrated to date, using either metallic or all-dielectric structures, have required a large driving voltage exceeding 10~V to achieve sufficient modulation swing. 
For energy-efficient operation and dense integration of multi-pixel modulators with a high-speed complementary metal-oxide-semiconductor (CMOS) driver circuit, it is imperative to reduce the driving voltage, ideally to around 1~V or smaller.

In this work, we demonstrate an efficient EO metasurface modulator operating at a subvolt driving voltage for the first time to our knowledge. The device is based on a silicon-organic-hybrid (SOH) metasurface incorporating a dimerized grating nanostructure.
This unique geometry enables the coupling of normally incident free-space light to a quasi-bound state in the continuum (\textit{q}BIC), which is a high-\textit{Q} resonating mode confined between the silicon bars.
Consequently, both the optical mode and the applied electrical fields are co-localized inside a narrow silicon slot region embedded with an OEO material, boosting the modulation efficiency.
The fabricated device with a 60-{\textmu}m squared size achieves 
a substantial reflectance modulation of $\Delta R = 0.63$ at an applied voltage of 3.3~V. 
The device is then used to generate 50-Mbps non-return-to-zero (NRZ) and 100-Mbps 4-level pulse amplitude modulation (PAM4) optical signals with peak-to-peak driving voltages ($V_\text{pp}$) of only 0.2~V and 0.8~V, respectively.
Furthermore, we effectively reduce the active modulator size down to a 10-{\textmu}m square by integrating distributed Bragg reflector (DBR) sections. Thanks to the reduced capacitance, the EO bandwidth is extended to 0.38~GHz, which is the highest reported to date among all-dielectric active-metasurface modulators. 
Finally, data modulations of 1.0-Gbps NRZ and 1.6-Gbps PAM4 signals are demonstrated with $V_\text{pp}=$ 1~V.

Compared to previously demonstrated EO metasurface modulators, including both nanometallic and all-dielectric structures, our 60-{\textmu}m-squared-size device exhibits a record-low driving voltage of 2.5~V to fully shift the resonant wavelength, while offering a substantial modulation swing of $\Delta R=0.63$.
Furthermore, the 10-{\textmu}m-squared-size device achieves, for the first time to our knowledge, high-speed modulation of pseudo-random data patterns exceeding Gbps using an active metasurface of any kind.
The subvolt operation demonstrated in this work enables the use of amplifier-free CMOS driving circuits, paving the way for the realization of high-speed energy-efficient active metasurface devices.

\section*{Concept of SOH metasurface modulator with dimerized grating}

\begin{figure*}
\centering\includegraphics{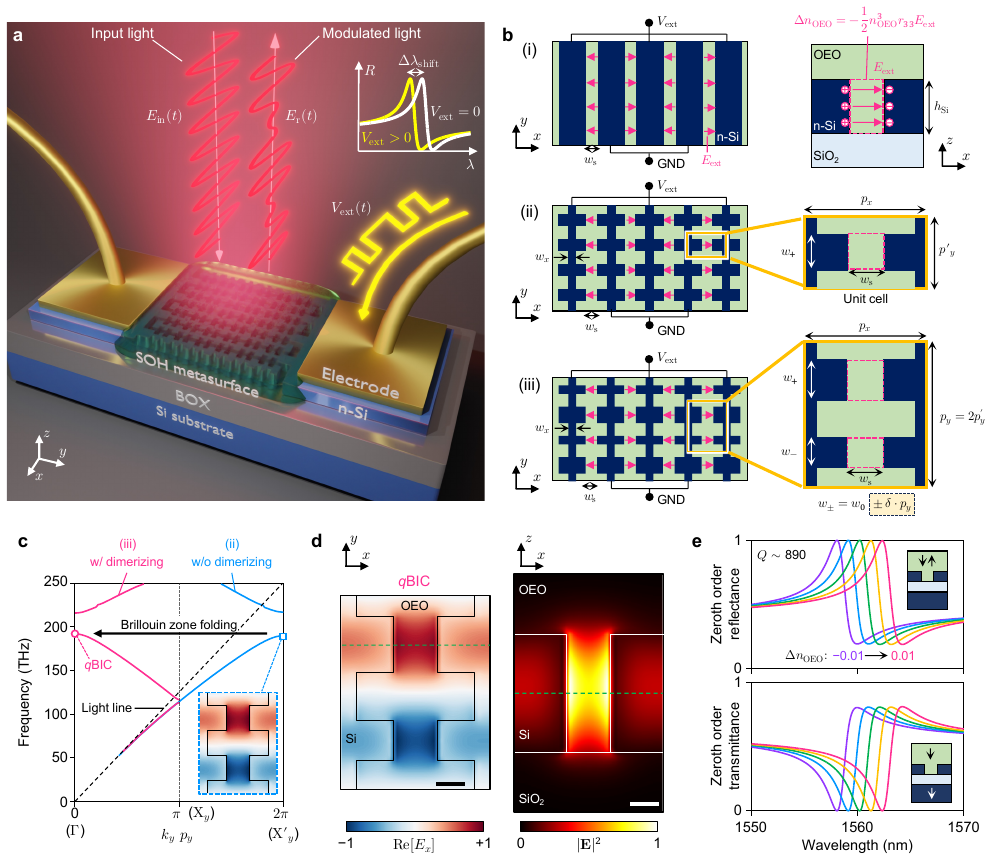}
\caption{
\textbf{Metasurface modulator with an all-dielectric SOH dimerized resonator.} 
\textbf{a},~Schematic illustration of our SOH free-space modulator. The intensity of the reflected light $|E_r(t)|^2$ can be modulated at high speed via an external voltage $V_\text{ext}(t)$. 
\textbf{b},~Top views of three types of SOH gratings: (i) slot waveguide array, (ii) single-periodic grating, (iii) dimerized grating. An external voltage $V_\text{ext}$ between silicon interdigitated bars induces a refractive index change $\Delta n_\text{OEO}$ of the embedded OEO material (right-top panel). The regions enclosed by the pink dotted lines represent the OEO material within the narrow slot region, whose refractive index changes largely.
In (iii), to break the symmetry, the silicon rod widths $w_\pm$ are perturbed by $\pm \delta\cdot p_y$.
\textbf{c},~Photonic band diagrams of single-periodic grating (blue) and dimerized grating (pink) nanostructures. Here, $k_y$ is swept while $k_x$ is fixed to 0. 
Through the Brillouin zone folding incurred by the period-doubling perturbation (dimerization), the bound mode at the X$'_y$ point (blue square), which is located below the light line, is transferred to the \textit{q}BIC mode at the $\rm{\Gamma}$ point (pink circle), which is accessible by normal-incident free-space light.
The inset shows the simulated electric field distribution Re$[E_x(x,y)]$ of the bound mode at the X$'_y$ point. 
\textbf{d},~Electric field distributions Re$[E_x(x,y)]$ and $|\mathbf{E}(x,z)|^2$ of the \textit{q}BIC mode at the resonant wavelength of 1561~nm, which are well confined inside the narrow slot regions.
The green dotted line in each panel represents the cross-sectional position of the mode plotted in the other panel. 
The scale bars are 100~nm.
\textbf{e},~Simulated reflection and transmission spectra of $x$-polarized normally incident light as the refractive index of the OEO material in the slot region (within the pink dotted lines in \textbf{b}) changes by $\Delta n_\text{OEO}$.
In \textbf{c}-\textbf{e}, we assume $h_\text{Si}=400$~nm, $p_x=500$~nm, $p_y=650$~nm, $w_0=162.5$~nm, $w_s=150$~nm, $w_x=100$~nm, and $\delta=0.04$ (for the dimerized case).
}
\label{concept}
\end{figure*}

\noindent
Our metasurface modulator consists of a two-dimensional silicon grating nanostructure with an embedded OEO material (Fig.~\ref{concept}a).
The entire device can be fabricated through the mature silicon photonic process combined with a simple and low-cost spin-coating deposition of the OEO material.
Here, n-doped silicon (n-Si) bars function as both optically transparent resonators and highly conductive interdigitated electrodes \cite{Kosugi2016-yg}. 
Thus, an externally applied voltage $V_\text{ext}$ provides a strong electric field $E_\text{ext}$ within the narrow slot region between the silicon bars.
Due to the large Pockels coefficient of the OEO material, which can be as large as $r_{33}>200$~pm/V \cite{Jin2014-qn}, its refractive index changes by $\Delta n_\text{OEO}=-\frac{1}{2}n_\text{OEO}^3 r_{33} E_\text{ext}$, where $n_\text{OEO}$ is its initial refractive index.
This induces a wavelength shift $\Delta \lambda_\text{shift}$ of the resonant mode, resulting in intensity modulation of the reflected and transmitted light when $x$-polarized light near the resonant wavelength is normally incident to the device (Fig.~\ref{concept}a, inset).

Figure~\ref{concept}b describes our strategy to efficiently trap the normally incident light and couple it to a high-\textit{Q} resonant mode inside the narrow slot region.
The simplest structure is a periodic SOH waveguide array (Fig.~\ref{concept}b-(i)).
Such structure generally enables strong light confinement inside the narrow slot region embedded with the OEO material, which has been utilized to demonstrate various waveguide-based high-efficiency modulators \cite{Koos2009-tx, Kieninger2018-nc}.
Unfortunately, however, this slot waveguide mode cannot be directly excited by a surface-normal incident light.
Moreover, the mode is not confined in the propagation direction (along the $y$ axis), so that high-\textit{Q} resonance cannot be obtained by itself.
To cope with this issue, we could employ the so-called high-contrast-grating (HCG) modes, which exhibit high-$Q$ lateral resonances and can be excited by normally incident light \cite{Kosugi2016-yg, Ogasawara2019-cp}.
This approach, however, lacks flexibility in design and has limited efficiency.

In this work, we overcome these limitations by the following two-step structure engineering (Fig.~\ref{concept}b-(ii) and -(iii)).
We first introduce periodic notches in the waveguiding direction ($y$) (Fig.~\ref{concept}b-(ii)). 
The period $p'_y$ of these notches is set approximately to a half wavelength of the slot waveguide mode.
Due to the in-plane Bragg reflection by the periodic structure, the fundamental mode at the X$'_y$ point ($k_y=\pi/p'_y$) in the photonic band diagram (Fig.~\ref{concept}c) has a standing-wave field distribution, which is localized in the $y$ direction and strongly confined within the narrow slot region (Fig.~\ref{concept}c, inset).
With sufficiently deep notches, we can induce a large band bending so that the group velocity $\mathrm{d}\omega/\mathrm{d} k_y$ around the X$'_y$ point is kept small.
As a result of this slow-light effect, the entire device can be reduced in the $y$ direction without incurring lateral leakage.
However, similar to the SOH waveguide mode, this mode is still located below the light line, so it cannot be excited by surface-normal light.

To enable coupling to normally incident light, we introduce an additional perturbation to the structure by varying the rod width of the silicon grating bars in every other period (Fig.~\ref{concept}b-(iii)).
The resulting structure has two silicon rods having slightly different widths of $w_\pm=w_0 \pm \delta \cdot p_y$ inside the doubled period of $p_y = 2 p'_y$, where $\delta$ is a dimensionless perturbation parameter.
As a result of this period doubling, the photonic band is folded as shown in Fig.~\ref{concept}c, so that the bound mode initially at the X$'_y$ point is transferred to the $\Gamma$ point ($k_y=0$, pink circle) \cite{Overvig2018-as}.
In the context of BIC, this mode at the $\Gamma$ point, which weakly couples to the normal-incident light, 
is referred to as a \textit{q}BIC mode \cite{Hsu2016-mi}.
It is important that the radiation \textit{Q} factor of this mode can be expressed as $Q_\text{rad}\propto \delta^{-2}$, so that the coupling to the surface-normal incident light can be controlled precisely by tuning $\delta$ \cite{Koshelev2018-zn}. 
Thus, we can realize both strong lateral confinement by introducing deep notches and sufficiently high-\textit{Q} resonance by adjusting $\delta$. 
Moreover, the resonant wavelength $\lambda_\text{res}$ can also be controlled independently by tuning the period $p_y$.

Figure~\ref{concept}d shows the simulated field distribution of the \textit{q}BIC mode at the $\Gamma$ point when $\delta = 0.04$ (see Methods for the details of the simulation).
We can confirm that the field is confined inside the narrow slot region with $w_\text{s}=150$~nm, similar to the bound mode (Fig.~\ref{concept}c, inset).
The confinement factor of the \textit{q}BIC mode is as large as $\Gamma_\text{OEO}=0.25$ (see Methods for definition).
Based on the perturbation theory \cite{Joannopoulos2008-qj}, the resonant wavelength shift $\Delta \lambda_\text{shift}$ is given by $\Delta\lambda_\text{shift}/\lambda_\text{res} \sim (\Delta n_\text{OEO} / n_\text{OEO})\Gamma_\text{OEO}\propto -\Gamma_\text{OEO}V_\text{ext}/w_\text{s}$.
Due to the large $\Gamma_\text{OEO}$ and small $w_s$, we can obtain large $\Delta \lambda_\text{shift}$ with small $V_\text{ext}$.
Moreover, unlike the bound mode at the X$'_y$ point, this \textit{q}BIC mode at the $\Gamma$ point can be coupled to the surface-normal free-space light.
Figure~\ref{concept}e shows the simulated reflection and transmission spectra. 
As we change the refractive index of the OEO material in the slot region by $\Delta n_\text{OEO}$, the resonant wavelength shifts linearly without increasing the insertion loss.

\begin{figure*}
\centering\includegraphics{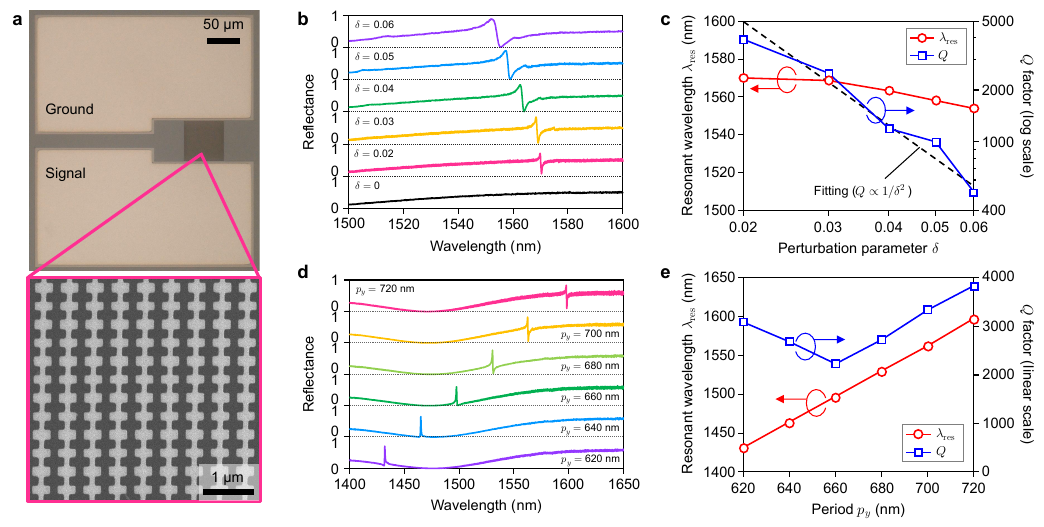}
\caption{
\textbf{Passive characterization of the fabricated devices.} 
\textbf{a}, Optical (top) and SEM (bottom) images of a fabricated device with a 60-{\textmu}m-squared size before spin-coating the OEO material.
\textbf{b}, Measured reflectance spectra with various dimerization factors $\delta$ ($p_y=710$~nm).
\textbf{c}, Resonant wavelengths and $Q$ factors extracted from the measured spectra in \textbf{b} as a function of $\delta$. A fitting curve with $Q\propto 1/\delta^2$ is also plotted.
\textbf{d}, Measured reflectance spectra with linearly changed periods $p_y$ ($\delta=0.02$).
\textbf{e}, Resonant wavelengths and $Q$ factors extracted from the measured spectra in \textbf{d} as a function of $p_y$. 
}
\label{passive}
\end{figure*}

We should note that our device is fundamentally different from SOH metasurface modulators using guided-mode resonance (GMR) \cite{Zheng2024-xh}. 
In the GMR scheme, a periodic perturbation is introduced to the SOH waveguide along the $y$ direction to form a grating with a period twice that of the case shown in Fig.~\ref{concept}b-(ii).
Since this perturbation also determines the coupling of the mode to free-space light, only shallow grating can be introduced to maintain the \textit{Q} factor high.
As a result, the lateral confinement of the mode along the $y$ direction is limited, which inevitably requires a longer device \cite{Overvig2018-as}.
In contrast, our dimerized metasurface nanostructure allows independent tuning of the band bending to achieve strong modal confinement in the $y$ direction (by the deep notches) and optimal coupling to surface-normal light (by optimized $\delta$).
Consequently, high-\textit{Q} resonance can be achieved with a significantly smaller device size, which is essential for high-speed modulation and high-density integration.
 (See Supplementary Section~1 for a detailed discussion and comparison.)

\begin{figure*}
\centering\includegraphics{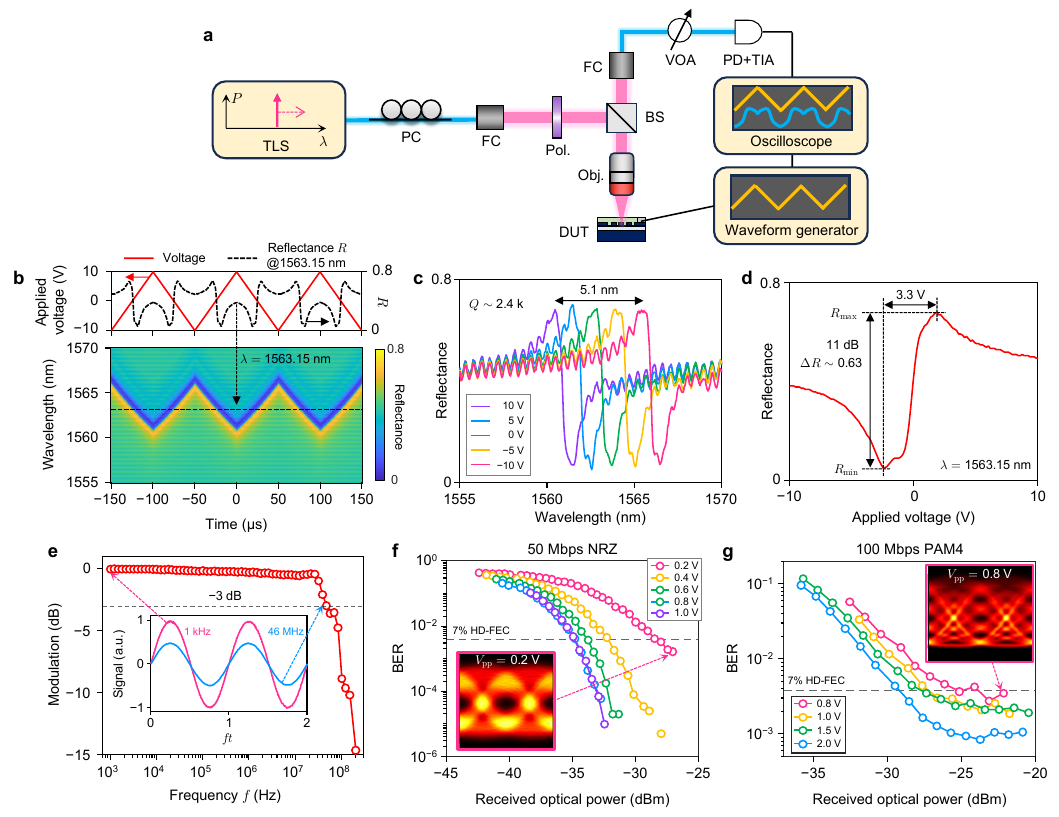}
\caption{
\textbf{Active modulation of free-space light.} 
\textbf{a}, Measurement setup. The electrical signals from the waveform generator were set to triangle, sinusoidal, and data modulation signals for the measurements in \textbf{b}-\textbf{d}, \textbf{e}, and \textbf{g}-\textbf{f}, respectively. 
TLS: tunable-wavelength laser source. PC: polarization controller. FC: fiber collimator. Pol.: polarizer. BS: beam splitter. Obj.: objective lens. DUT: device under test. VOA: variable optical attenuator. PD: photodetector. TIA: trans-impedance amplifier.
\textbf{b}, Reflection waveforms (bottom) when the triangle wave voltage (red line in top panel) is applied in each wavelength. 
\textbf{c}, Reflectance specta extracted from \textbf{b} with applied voltages ranging from $-10$ to 10~V.
\textbf{d}, Reflectance modulation as a function of the applied voltage at a wavelength of 1563.15~nm.
\textbf{e}, Modulation depth of the measured reflectance as a function of the frequency $f$ in the sinusoidal modulation signal. The inset shows the measured PD signals at $f=1$~kHz and 46~MHz.
\textbf{f},\textbf{g}, BER curves of 50-Mbps NRZ (\textbf{f}) and 100-Mbps PAM4 (\textbf{g}) signals as a function of the received optical power. The insets show the measured eye diagrams.
}
\label{active}
\end{figure*}

\section*{Demonstration of SOH metasurface modulator}

\noindent
The SOH metasurface modulators were fabricated using a silicon-on-insulator (SOI) substrate with a 400-nm-thick silicon layer on a 2-{\textmu}m-thick buried oxide (BOX) layer (see Methods for the detailed fabrication process).
As the OEO material, we employed JRD1 (NLM Photonics) \cite{Jin2014-qn}.
We employed the same grating parameters as those assumed in Fig.~\ref{concept}c-e, 
except for $p_y$ and $\delta$, which were varied on each device.
The area of the grating region in each device was set to $60\times60$~{\textmu}m$^2$.
Figure~\ref{passive}a shows optical microscope and scanning electron microscope (SEM) images of a fabricated device with $p_y=710$~nm and $\delta = 0.02$.

Figure~\ref{passive}b shows the reflection spectra of various devices with $p_y=710$~nm and different values of $\delta$, measured before poling the OEO material. We can confirm that a resonance peak, which is absent in the un-dimerized device ($\delta=0$), emerges as we introduce dimerization ($\delta>0$). The resonant wavelength and the $Q$ factor for each case are plotted in Fig.~\ref{passive}c as a function of $\delta$. The $Q$ factor clearly follows the theoretical curve ($Q\propto 1/\delta^2$) with a relatively small change ($<$15~nm) in the resonant wavelength. 
Figure~\ref{passive}d shows the measured reflection spectra of different devices with $\delta=0.02$ and $p_y$ ranging from 620 to 720~nm. The resonant wavelength $\lambda_\text{res}$ and the $Q$ factor are plotted in Fig.~\ref{passive}e as a function of $p_y$. Now, we can see that $\lambda_\text{res}$ changes linearly with $p_y$, while the $Q$ factor is kept high at more than 2,000 in all cases.
We can therefore control the $Q$ factor and $\lambda_\text{res}$ independently by tuning $\delta$ and $p_y$, respectively.
We should also note that the maximum reflection at the resonant peak is greater than 0.6 for all the cases shown in Fig.~\ref{passive}b,d.
From these results, we select the device with $\delta=0.02$ and $p_y = 700$~nm in the following measurements to operate at a wavelength around 1560~nm.

Figure~\ref{active}a shows our setup to characterize the active modulation properties after poling the OEO material. The $x$-polarized light from a tunable-wavelength laser source (TLS) was normally incident to the device, which was electrically driven by a waveform generator through a DC probe. The reflected light from the device was converted to an electrical signal by a photodetector (PD), which was then observed by a real-time oscilloscope. (See Methods for detailed explanation.)

Figure~\ref{active}b shows the observed waveform of the reflected optical power at each wavelength, swept from 1555 to 1570~nm, when a 10-kHz triangular waveform voltage is applied to the device.
From this result, the reflection spectra under different applied voltages are obtained (Fig.~\ref{active}c). 
We can see that the $Q$ factor remains nearly constant at approximately 2,400 across all cases, whereas the resonant wavelength $\lambda_\text{res}$ shifts linearly with the applied voltage in agreement with the theory (Fig.~\ref{concept}e).
The measured wavelength shift is as large as $\Delta\lambda_\text{shift}=5.1$~nm at $\pm10$~V, corresponding to an efficiency of $S\equiv\Delta \lambda_\text{shift}/\Delta V=0.26$~nm/V.
Based on the simulated confinement factor $\Gamma_\text{OEO} = 0.25$, the in-device EO coefficient $r_{33}$ of the OEO material is estimated to be 58~pm/V.

To quantify the modulation efficiency of our device, we define $V_\text{req}\equiv \Delta \lambda_\text{res}/S$, where $\Delta \lambda_\text{res}\equiv \lambda_\text{res}/Q$ is the linewidth of the resonance.
$V_\text{req}$ denotes the voltage required to shift the resonant wavelength by an amount equal to its linewidth, which corresponds to the necessary condition to achieve sufficient modulation swing.
From the results shown in Fig.~\ref{active}c, we derive $V_\text{req}=2.5~\text{V}$, which, to the best of our knowledge, is the lowest value reported for high-speed active metasurfaces (see Extended Data Table~\ref{tab:benchmark}). 
Figure~\ref{active}d shows the measured reflectance as a function of the applied voltage at a wavelength of 1563.15~nm. 
Efficient modulation with $\Delta R = R_\text{max}-R_\text{min} = 0.63$ (where $R_\text{max}$ and $R_\text{min}$ are the maximum and minimum reflectances), an extinction ratio of 11~dB, and an optical insertion loss of 2~dB is achieved experimentally with a small voltage of 3.3~V.
Figure~\ref{active}e shows the EO frequency response of the device, which is measured by applying a sinusoidal electrical signal at different frequencies. The 3-dB bandwidth is derived to be 46~MHz.

Thanks to the ultrahigh modulation efficiency and low optical insertion loss that allows for a high optical signal-to-noise ratio (OSNR), our device can be used for high-speed data modulation by a random signal pattern.
Figures~\ref{active}f,g show the results of driving our device using $2^9-1$ pseudo-random binary sequence (PRBS) patterns.
A bit-error rate (BER) below the threshold of 7\% hard-decision forward error correction (HD-FEC) is obtained for 50-Mbps NRZ and 100-Mbps PAM4 signals with $V_\text{pp}$ of only 0.2~V and 0.8~V, respectively.
To the best of our knowledge, this is the first demonstration of pseudo-random data modulation using an EO metasurface.

\begin{figure*}
\centering\includegraphics{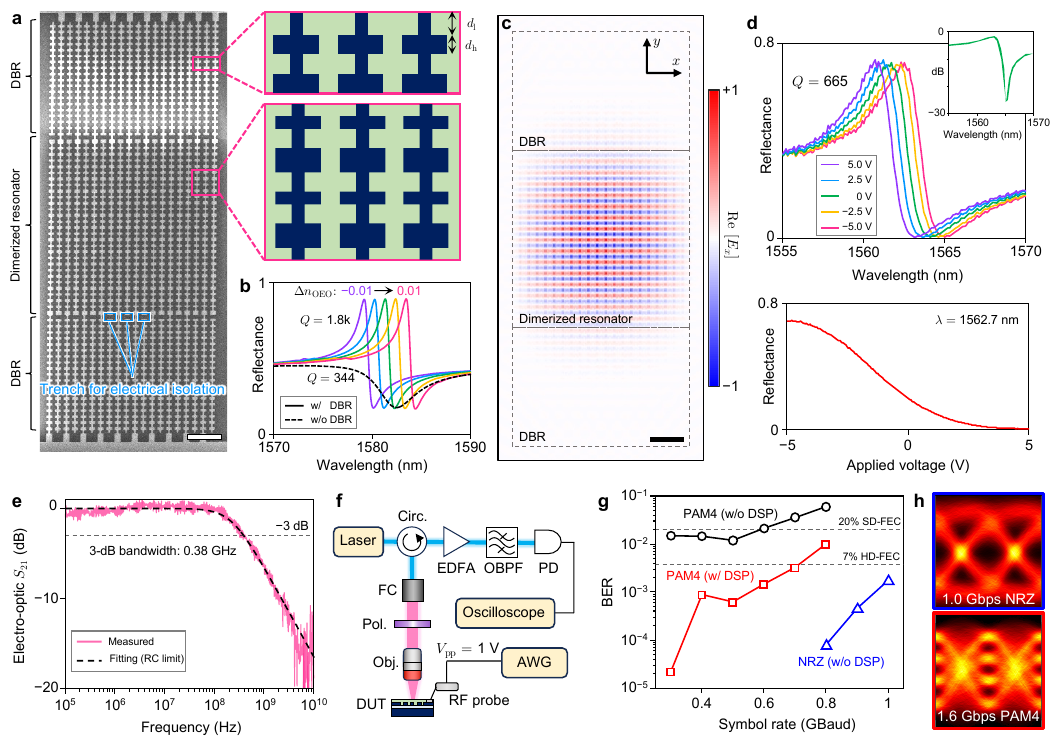}
\caption{
\textbf{High-speed modulation using an ultra-small DBR-integrated device.} 
\textbf{a}, SEM image of the fabricated device composed of a 10-\textmu{m}-squared dimerized resonator with 20-layer DBR patterns on both sides. Scale bar is 2~{\textmu}m. The insets show magnified schematics of the DBR and dimerized patterns.
\textbf{b}, Simulated reflection spectra as the refractive index of the OEO material within the slot changes. The case without the DBR sections is plotted by the dotted line for reference.
\textbf{c}, Simulated electric field distribution of the resonant mode. Scale bar is 2~{\textmu}m.
\textbf{d}, Top: Measured reflectance specta under different applied voltages from $-5$ to 5~V. The inset shows the measured reflection spectrum at 0~V in the dB scale. Bottom: Reflectance as a function of the applied voltage at a wavelength of 1562.7~nm. 
\textbf{e}, Measured EO frequency response. The fitting curve for the RC-limited case is plotted by the dotted line.
\textbf{f}, Setup for high-speed data modulation experiment.
\textbf{g}, Experimentally obtained BERs of NRZ and PAM4 signals modulated by $V_\text{pp} = 1$~V as a function of the symbol rate.
\textbf{h}, Observed eye patterns of 1.6-Gbps PAM4 signal with DSP-based equalization and 1.0-Gbps NRZ signal without DSP.
}
\label{mini}
\end{figure*}

\section*{Higher-speed device with DBR sections}

\noindent
For higher-speed modulation, it is essential to shrink the device size since the EO bandwidth of our modulator is RC-limited (see Supplementary Section 2 for detailed analysis).
To significantly reduce the length of the active grating region down to a 10-{\textmu}m range without degrading the $Q$ factor, we introduce DBR sections.
The fabricated device is shown in Fig.~\ref{mini}a. 
It consists of a 10-{\textmu}m-squared dimerized resonator with 
the similar grating parameters as those assumed in Fig.~\ref{concept}c-e (see Supplemental Table~1 for the full list of parameters), attached with DBR patterns on both sides of the grating.
To minimize parasitic capacitance, we inserted electrical isolation trenches between the dimerized grating resonator and DBR sections (blue boxes in Fig.~\ref{mini}a).
The DBR consists of narrow and wide slots (Fig.~\ref{mini}a, inset), each having a different effective refractive index of $n_\text{h}$ or $n_\text{l}$, where $n_\text{h}>n_\text{l}$. The length of each section was selected to $d_\text{h}=\lambda/4n_\text{h}=154$~nm and $d_\text{l}=\lambda/4n_\text{l}=198$~nm to satisfy the Bragg condition.

Figure~\ref{mini}b shows the simulated reflection spectrum under modulation (see Supplementary Fig.~3 for the simulated spectra of transmission and leakage loss).
In a case without the DBRs (dotted line), the $Q$ factor of the resonant mode is significantly degraded compared to the infinite-size case with the same design of the dimerized grating (Fig.~\ref{concept}e).
In contrast, by introducing DBRs, the optical field is effectively confined within the 10-{\textmu}m-squared dimerized resonator region (Fig.~\ref{mini}c). As the lateral leakage is suppressed, the high-$Q$ \textit{q}BIC mode is retained, and a large change in the reflection spectrum is obtained at the resonant wavelength. 

Figure~\ref{mini}d shows the experimentally measured reflection spectra under different applied voltages (top) and the reflectance plotted as a function of voltage at a wavelength of 1562.7~nm (bottom).
In the inset of Fig.~\ref{mini}d, we also plot the reflection spectrum in a logarithmic scale for the unbiased case, indicating that 20-dB extinction is obtained at the resonant wavelength.
From Fig.~\ref{mini}d, we derive the modulation efficiency and the required voltage to be $S=0.18$~nm/V and $V_\text{req}=11.2$~V, respectively.

Figure~\ref{mini}e presents the EO frequency response measured by a vector network analyzer (VNA), which nicely fits with the RC-limited model.
We obtain the 3-dB EO bandwidth of 0.38~GHz, which, to our knowledge, is the record-high value for all-dielectric active metasurfaces reported to date (see Extended Data Table~\ref{tab:benchmark}).
Finally, we performed a data transmission experiment at 1560.2~nm using the setup shown in Fig.~\ref{mini}f (see Methods for a detailed explanation).
The $V_\text{pp}$ of the modulation voltage was fixed to 1~V for all cases.
In Fig.~\ref{mini}g, we plot the obtained BER with increasing symbol rates for various modulation formats. For the PAM4 format, we compared the cases with and without applying the signal equalization based on digital signal processing (DSP) (see Methods for a detailed explanation). 
Figure~\ref{mini}h shows the observed eye patterns of modulated 1.6-Gbps PAM4 and 1.0-Gbps NRZ signals, where we successfully achieved BERs below the FEC threshold.

\begin{figure}
\centering\includegraphics{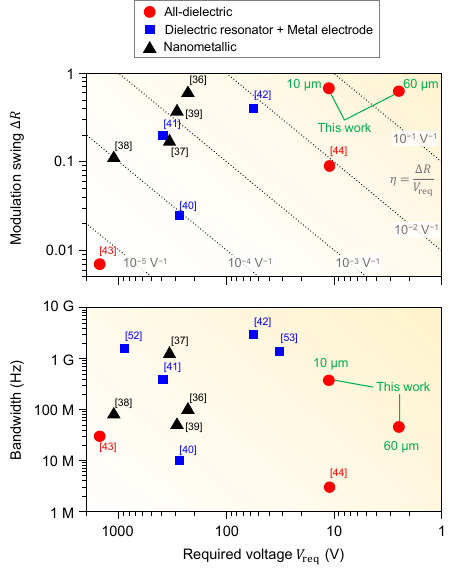}
\caption{
\textbf{Benchmark comparison of this work against previously reported EO metasurface modulators.} 
The modulation swing $\Delta R$ (top panel vertical axis), bandwidth (bottom panel vertical axis), and the required voltage $V_\text{req}$ to shift the resonant wavelength by an amount equal to its linewidth (horizontal axis) of experimentally demonstrated EO metasurface modulators in the literature.
Our presented device exhibits the lowest $V_\text{req}$ and the highest modulation efficiency $\eta=\Delta R/V_\text{req}$. The bandwidth is also the highest among the all-dielectric resonating devices. The numerical values are listed in Extended Data Table~\ref{tab:benchmark}.
}
\label{benchmark}
\end{figure}

\section*{Conclusions}
\noindent
We have demonstrated free-space metasurface-based optical modulators with greatly improved overall performance (i.e., modulation efficiency, insertion loss, and EO bandwidth) by employing a dimerized SOH grating nanostructure.
The unique grating geometry of our device allows for optimal coupling of surface-normal incident light to a \textit{q}BIC mode that is strongly confined inside the narrow silicon slot region, enabling highly efficient modulation.

Figure~\ref{benchmark} and Extended Data Table~\ref{tab:benchmark} compare the performance of our demonstrated devices with previously reported experimental work on EO metasurface modulators using both metallic and dielectric resonating structures.
The required voltage $V_\text{req}$ (defined as the voltage to shift the resonant wavelength by an amount equal to its linewidth) of our 60-{\textmu}m-squared-size device is as small as 2.5~V.
Combined with the large modulation swing (i.e., low insertion loss and large extinction ratio) of $\Delta R=0.63$, it exhibits the highest modulation efficiency of $\eta\equiv \Delta R / V_\text{req}=0.25$~V$^{-1}$, which is more than an order-of-magnitude improvement from previous demonstrations.
Using this device, we have successfully generated 50-Mbps NRZ and 100-Mbps PAM4 optical signals with driving voltages $V_\text{pp}$ of less than 1~V.
In addition, our 10-{\textmu}m-square device with DBR structures has an EO bandwidth of 0.38~GHz, which is the highest among all-dielectric active metasurface modulators demonstrated to date (Fig.~\ref{benchmark}, bottom).
Thanks to the high modulation efficiency and broad bandwidth, we have achieved transmission of 1.0-Gbps NRZ and 1.6-Gpbs PAM4 signals with $V_\text{pp}=1$~V.
To the best of our knowledge, this is the first experimental demonstration of high-speed optical modulation of pseudo-random signals with a bitrate exceeding Gbps using an active metasurface of any kind.

We expect that these metrics can be further improved as described below.
\begin{itemize}
    \item The estimated $r_{33}$ of 58~pm/V in our device is far below the previously reported value of 390~pm/V achieved by a waveguide-based SOH modulator using a similar OEO material \cite{Kieninger2018-nc}.
    Thus, by improving the fabrication process and optimizing the poling conditions, the driving voltage of our device should be decreased further.
    \item The measured EO bandwidth of 0.38~GHz of our 10-{\textmu}m-square-size device is one-order smaller than the theoretically expected bandwidth of $\sim$2.6~GHz (see Supplementary Fig.~2). From a separate measurement, we confirm that our device exhibits Schottky-like behavior with a large contact resistance of $>$1~k$\mathrm{\Omega}$.
    The bandwidth should thus be increased by improving the doping and metalization processes to achieve low-resistance ohmic contact. 
    \item To further enhance the modulation bandwidth, the electrical resistance of the interdigitated Si grating bars needs to be reduced. 
    We can achieve this by increasing the doping concentration of the silicon layer up to a level where the optical absorption in the doped silicon does not degrade the \textit{Q} factor of the optical resonance substantially. Furthermore, replacing Si with GaAs or InP can significantly decrease the electrical resistance due to the higher electron mobility with lower optical absorption \cite{Fukui2022-my}.
    Using highly doped n-InP, for example, we can reach $>$100~GHz bandwidth with a device dimension of $<$12~{\textmu}m (see Supplementary Fig.~2).
\end{itemize}

While we have focused on a single-pixel reflective intensity modulator in this work, our device can be readily extended to realize other functionalities depending on the application. Specifically, a transmissive intensity modulator, phase modulator, and two-dimensional multi-pixel spatial light modulator (SLM) should also be feasible through minor changes in the device structure.
The subvolt operation of our modulator demonstrated in this work allows the use of amplifier-free CMOS driver circuits, enabling energy-efficient and high-speed active metasurface devices for optical communication, sensing, and computing applications.

\bibliography{reference}
\clearpage
\footnotesize
\section*{Methods}

\subsection*{Optical simulation}
\noindent
In all simulations, we employed a three-dimensional finite-difference time-domain (FDTD) method (Ansys: Lumerical FDTD). Here, we set the refractive indices of silicon, silicon dioxide, and OEO material to 3.48, 1.444, and 1.84, respectively.
To obtain the band diagrams shown in Fig.~\ref{concept}c, the periodic and Bloch boundary conditions were employed at the $x$ and $y$ boundaries, respectively, and the Bloch wave vector $k_y$ was swept while $k_x$ was fixed to 0. Here, randomly arranged dipole sources were utilized to excite optical modes in the metasurface.
To avoid the excitation of irrelevant guided modes, the silicon substrate and air layers were eliminated, and infinitely thick OEO and BOX layers were assumed.
To obtain the results shown in Fig.~\ref{concept}d,e, we employed the periodic boundary condition at both the $x$ and $y$ boundaries and plane-wave incidence at a normal angle, as well as the infinite thickness of the OEO layer and silicon substrate.
For simplicity, the refractive index was assumed to change uniformly only in the narrow slot OEO regions inside the pink rectangles depicted in Fig.~\ref{concept}b.
In simulating the reduced-size device (Fig.~\ref{mini}b,c), all boundaries were set to perfectly matched layers (PMLs), and a Gaussian beam with a field diameter of 10~{\textmu}m was employed as an input source.
Here, a graphical processing unit (GPU) (NVIDIA GeForce RTX4090) was utilized to accelerate the simulation.

\subsection*{Definition and derivation of confinement factor}
\noindent
We assume that the applied electric field $E_\text{ext}$ is uniform and oriented in the $x$ direction inside the narrow slot region. We also ignore the Pockels effect in the other regions.
Under these assumptions, the confinement factor $\Gamma_\text{OEO}$ is expressed as \cite{Joannopoulos2008-qj, Benea-Chelmus2021-re, Zheng2024-xh}
\begin{equation}
    \Gamma_\text{OEO} = \frac{\iiint_{\text{OEO}} \varepsilon_\text{OEO}\left|E_x(\mathbf{r}) \right|^2 \mathrm{d} x \mathrm{d} y \mathrm{d} z}{\iiint_{\text{all}} \varepsilon(\mathbf{r})\left|\mathbf{E}(\mathbf{r})\right|^2 \mathrm{d} x \mathrm{d} y \mathrm{d} z},
\end{equation}
where $\varepsilon_\text{OEO}$ is the permittivity of the OEO material, $\varepsilon(\mathbf{r})$ is the permittivity at the position $\mathbf{r}$, and $\mathbf{E}=(E_x, E_y, E_z)^t$ represents the electric field vector of the optical mode. 
Here, the integral range of the denominator is the entire space while that of the numerator is only inside the narrow slot region, enclosed by pink dotted lines in Fig.~\ref{concept}b.
$\Gamma_\text{OEO}$, therefore, describes the ratio of the electric field energy of the optical mode inside the narrow slot OEO region.

\subsection*{Device fabrication}
\noindent
The fabrication flow is shown in Supplementary Fig.~4.
The SOH modulator was fabricated on an SOI substrate with silicon, BOX, and substrate thicknesses of $h_\text{Si}=400$~nm, 2~{\textmu}m, and 725~{\textmu}m. The doping concentration of the top n-type silicon layer was $\sim1\times10^{18}$~cm$^{-3}$.
The positive electron-beam resist (ZEON: ZEP520A-7) was first spin-coated on a $1.3\times1.3$~cm$^2$ chip. The metasurface pattern was written on the resist using an electron beam (EB) writer (ADVANTEST: F7000S), and the pattern was developed by a resist developer (ZEON: ZED-N50). Then, the pattern was transferred to the silicon layer using reactive-ion etching (RIE) with SF$_6$ and C$_4$F$_8$ gas (SPTS: MUC-21 ASE-Pegasus), known as the Bosch process, followed by O$_2$ ashing process.
The contact and pad patterns were defined on a spin-coated photoresist using a laser writer (HEIDELBERG: DWL66+).
The contact and pad metals were formed by a liftoff process after the deposition of aluminum layers ($\sim$400~nm) using an EB evaporator, which were then annealed at 400~$^\circ$C to reduce the contact resistance.
Finally, the OEO material was spin-coated. In this work, we used JRD1 (NLM Photonics) \cite{Jin2014-qn} as the OEO material.
Before measuring the modulation characteristics of the fabricated device, JRD1 was poled at 100~V/{\textmu}m under $\sim$85~$^\circ$C.

\subsection*{Measurement setup}
\subsubsection*{Broadband reflectance}
\noindent
To measure the passive reflectance over a broad wavelength range as shown in Fig.~\ref{passive}b,d, we used a wideband source (YSL Photonics: SC-5), an optical spectrum analyzer (OSA) (Ando: AQ6331), and the same free-space optics as shown in Fig.~\ref{active}a.
The light from the wideband source was emitted in free space using a fiber collimator (FC). 
The polarization state was aligned to the polarizer using a fiber-based polarization controller (PC). The $x$-polarized light reflected by a beam splitter (BS) was focused onto the device through an objective lens (Mitsutoyo: M Plan Apo NIR 5×) with a spot size of $\sim$30~{\textmu}m.
The reflected light transmitting through the BS was coupled to a fiber via another FC and then measured by the OSA with a spectral resolution of 0.1~nm. 
To calibrate the optical transmission loss of our measurement system, we employed a reference mirror (i.e., large-area aluminum pattern) fabricated on the same chip.
By normalizing the measured optical power from our device to that from the reference mirror, reflectance from our device was obtained.

\subsubsection*{Active modulation using a 60-{\textmu}m-square-size device}
\noindent
Figure~\ref{active}a shows the optical setup to characterize the active modulation. The free-space optical components were the same as those used to measure the passive reflectance.
Here, we used a TLS (Santec: TSL550) as the light source. An electrical signal from a waveform generator (NF: WF1968) was applied to the fabricated device through a DC probe. The reflected light was detected by a PD with an integrated trans-impedance amplifier (TIA) (New Focus: 1811-FC) and observed by a real-time oscilloscope (Tektronix: MDO4104C). 

To characterize the spectrum shift as shown in Fig.~\ref{active}b-d, the electrical signal was set to a 10-kHz triangular waveform with $V_\text{pp} = 20$~V. For each laser wavelength swept from 1555 to 1570~nm, we recorded the waveform of reflected optical power using the oscilloscope, which was averaged 128 times. By normalizing the results to that measured for a reference mirror, the reflectance waveforms were obtained as shown in Fig.~\ref{active}b.
For the measurement of the frequency response shown in Fig.~\ref{active}e, we switched the applied electrical signal to a sinusoidal waveform with $V_\text{pp}=0.5$~V. The number of averaging in the oscilloscope was set to 512 to increase the accuracy.

For data modulation, we applied a raised-cosine signal with a roll-off factor of 1 to the device. The received optical power before the PD was controlled by a variable optical attenuator (VOA). The detected signal by the PD was sampled at 250~MS/s in real-time by the oscilloscope. After the timing recovery and resampling of the waveform, the BER was calculated offline. 

\subsubsection*{Characterization of high-speed 10-{\textmu}m-square-size device}
\noindent
To characterize the reduced-size device, we built another measurement system as shown in Fig.~\ref{mini}f. The input light is emitted in free space and focused on the device with $x$ polarization. From the focal lengths of the FC (Thorlabs: F810APC-1550) and objective lens (Mitsutoyo: M Plan Apo NIR 10×), the spot size was calculated to be 5.6~{\textmu}m. The reflected light was coupled to the fiber and routed to the output port of a circulator.

For the passive reflection measurement (Fig.~\ref{mini}d, inset), we used amplified spontaneous emission (ASE) light from an erbium-doped fiber amplifier (EDFA) as a light source, and we measured the reflected light using an OSA (Finisar: WaveAnalyzer 1500S).
To obtain the spectrum shift shown in Fig.~\ref{mini}d, we switched the light source to the TLS and applied a 10~kHz triangular waveform as described in the previous section.
Here, we employed a ground-signal (GS) radio-frequency (RF) probe to enable higher-speed operation.
For the frequency response measurement, we employed a VNA (Anritsu: MS4640B) connected to the device (port 1) and a calibrated high-speed PD (port 2) (Anritsu: MN4765B).
For the data modulation experiment, we employed the setup shown in Fig.~\ref{mini}f.
The fabricated modulator was driven with a raised-cosine signal (roll-off factor = 1, $V_\text{pp}= 1$~V) from an arbitrary waveform generator (AWG) (Tektronix: AWG7102). The reflected light was amplified by an EDFA, filtered by an optical bandpass filter (OBPF), and detected by a high-speed PD (u$^2$t: XPDV2020R) with input power of $\sim$10~dBm. The detected signal was observed by a real-time oscilloscope.
We compared two cases with and without DSP-based signal equalization. In the former case, the recorded signal was resampled and demodulated with adaptive finite impulse response (FIR) filters with 11 taps.
In the latter cases, the recorded signal was time-recovered and resampled.
Finally, we derived the BER.

\section*{Acknowledgements}
\noindent
The authors acknowledge Y. Nakano and A. Otomo for their support and fruitful discussions.
G.S. thanks K. Saito for his help in building the measurement system.
This work was supported by the Japan Society for the Promotion of Science (JSPS), Grant Number JP23H05444, JP24KJ0557, and from the commissioned research by National Institute of Information and Communications Technology (NICT), Japan, Grant Number JPJ012368C0360.
The device was fabricated in part at Takeda Cleanroom with help of Nanofabrication Platform Center of School of Engineering, the University of Tokyo, Japan, supported by ``Advanced Research Infrastructure for Materials and Nanotechnology in Japan''
of the Ministry of Education, Culture, Sports, Science and Technology (MEXT), Grant Number JPMXP1224UT1115.
G.S. acknowledges financial support from JSPS.

\section*{Author contributions}
\noindent
G.S. and T.T. conceived the device concept and experiments.
G.S. performed the metasurface design, simulation, fabrication, and experiments.
K.A. developed the OEO process, constructed the poling setup, and helped the poling process.
S.K. assisted with numerical simulation.
Y.T. assisted with device fabrication.
G.S. and T.T. wrote the manuscript with inputs from all authors.
T.T. supervised the project.

\section*{Competing interests}
\noindent
G.S. and T.T. are listed as inventors in a patent application related to this work, filed by the University of Tokyo.

\clearpage

\setcounter{figure}{0}
\setcounter{table}{0}
\renewcommand{\figurename}{Extended Data Fig.}
\renewcommand{\tablename}{Extended Data Table}

\begin{table*}
\centering
\caption{\textbf{Comparison of modulation metrics in active EO metasurfaces.} 
Req. vol.:~Required voltage, Mod.:~Modulation, LN:~LiNbO$_3$, BTO:~BaTiO$_3$.}
\label{tab:benchmark}
\footnotesize
\begin{threeparttable}
\begin{tabular}{@{}cccccccccc@{}}
\toprule
\begin{tabular}{c}Active\\material\end{tabular} & \begin{tabular}{c}Resonator\\type\end{tabular} & 
\begin{tabular}{c}Electrode\\material\end{tabular} & \begin{tabular}{c}Wavelength\\(nm)\end{tabular} & $Q$ factor& \begin{tabular}{c}Req.~vol.\\ $V_{\text{req}}$~(V)\end{tabular} & \begin{tabular}{c}Mod.~swing\\$\Delta R$\end{tabular} & \begin{tabular}{c}Efficiency\\ \textbf{$\eta$} (V$^{-1}$)\end{tabular} & \begin{tabular}{c}Bandwidth\\(MHz)\end{tabular} & References \\ \midrule
OEO & Dielectric & Dielectric & 1560 & \textbf{2400} & \textbf{2.5} & \textbf{0.63} & \textbf{0.25} & 46 & This work \\
OEO & Dielectric & Dielectric & 1560 & 655 & 11.2 & \textbf{0.68} & 0.061 & \textbf{380} & This work \\
OEO & Nanometallic & Metal & 1280 & 200$^*$ & 229 & 0.6$^*$ & 0.0026 & 100 & \cite{Sun2022-wt} \\
OEO & Nanometallic & Metal & 1650 & 113 & 339 & 0.17 & 0.0005 & 1250 & \cite{Zhang2023-gb} \\
OEO & Nanometallic & Metal & 1290 & 77 & 1117 & 0.11 & 0.000098 & 80$^*$ & \cite{Zhang2023-de} \\
OEO & Nanometallic & Metal & 1390 & 70$^*$ & 289 & 0.37 & 0.0013 & 50 & \cite{Benea-Chelmus2021-re} \\
OEO & Dielectric & Metal & 1240 & 145 & 272 & 0.025$^*$ & 0.000092 & 10 & \cite{Sun2021-se} \\
OEO & Dielectric & Metal & 1310 & 153 & 389 & 0.2$^*$ & 0.00051 & 400 & \cite{Sun2022-as} \\
OEO & Dielectric & Metal & 1540 & 550 & 56 & 0.4 & 0.0071 & 3000 & \cite{Benea-Chelmus2022-di} \\
OEO & Dielectric & Dielectric & 1510 & 75 & 1510$^*$ & 0.007$^*$ & 0.0000046 & 30 & \cite{Ogasawara2019-cp} \\
OEO & Dielectric & Dielectric$^\dagger$ & 1510 & 1310 & 11 & 0.09$^*$ & 0.0082 & 3 & \cite{Zheng2024-xh} \\
LN & Nanometallic & Metal & 950 & 15 & N/A & 0.008 & N/A & 8 & \cite{Damgaard-Carstensen2022-yl} \\
LN & Nanometallic & Metal & 900 & 30 & N/A & 0.013 & N/A & 13.5 & \cite{Damgaard-Carstensen2023-ag} \\
LN & Nanometallic & Metal & 1550 & 180 & N/A & 0.35 & N/A & 125 & \cite{Damgaard-Carstensen2024-rh} \\
LN & Nanometallic & Metal & 770 & 129 & N/A & 0.000054 & N/A & 2.5 & \cite{Weigand2021-lu} \\ 
LN & Dielectric & Metal & 1550 & 1400 & 886 & N/A & N/A & 1600 & \cite{Trajtenberg-Mills2024-sx} \\
LN & Dielectric & Metal & 1550 & 8600 & 32 & N/A & N/A & 1400 & \cite{Di-Francescantonio2024-cu} \\
LN & Dielectric & Metal & 1570 & 50$^*$ & N/A & N/A & N/A & 83 & \cite{Ju2024-qc} \\
BTO & Nanometallic & Dielectric & 630 & 200 & N/A & N/A & N/A & 2 & \cite{Weigand2024-vi} \\
\bottomrule
\end{tabular}

\begin{tablenotes}
\footnotesize
\item[*] Extracted from figures.
\item[\dag] Au is also used but not required in principle.
\end{tablenotes}
\end{threeparttable}

\end{table*}

\end{document}